\documentclass[aps,pra,reprint,showpacs,superscriptaddress]{revtex4-1}
\usepackage{amsmath}
\usepackage{graphicx}
\usepackage{bm}
\begin{document}
\title{All-electrically reading out and initializing topological qubits with quantum dots}
\author{Wei Chen}
\email{pchenweis@gmail.com}
\affiliation{National Laboratory of Solid State Microstructures and Department of Physics, Nanjing University, Nanjing 210093, China}
\affiliation{Department of Physics and Center of Theoretical and Computational Physics, The University of Hong Kong, Pokfulam Road, Hong Kong, China}
\author{Zheng-Yuan Xue}
\affiliation{Department of Physics and Center of Theoretical and Computational Physics, The University of Hong Kong, Pokfulam Road, Hong Kong, China}
\affiliation{Laboratory of Quantum Information Technology, and School of Physics and Telecommunication Engineering, South China Normal University, Guangzhou 510006, China}
\author{Z. D. Wang}
\email{zwang@hku.hk}
\affiliation{Department of Physics and Center of Theoretical and Computational Physics, The University of Hong Kong, Pokfulam Road, Hong Kong, China}
\author{R. Shen}
\email{shen@nju.edu.cn}
\affiliation{National Laboratory of Solid State Microstructures and Department of Physics, Nanjing University, Nanjing 210093, China}
\author{D. Y. Xing}
\affiliation{National Laboratory of Solid State Microstructures and Department of Physics, Nanjing University, Nanjing 210093, China}
\begin{abstract}
We analyze the reading and initialization of a topological qubit encoded by Majorana fermions in one-dimensional semiconducting nanowires, weakly coupled to a single level quantum dot (QD). It is shown that when the Majorana fermions are fused by tuning gate voltage, the topological qubit can be read out directly through the occupation of the QD in an energy window. The initialization of the qubit can also be realized via adjusting the gate voltage on the QD, with the total fermion parity conserved. As a result, both reading and initialization processes can be achieved in an all-electrical way.
\end{abstract}
\maketitle

Topological computation based on Majorana fermions (MFs) has attracted much interest recent years \cite{Sarma}. By nonlocally encoding quantum information in pairs of separated MFs, most types of decoherence caused by local perturbations may be successfully avoided \cite{Kitaev}. Besides immune to decoherence, the MFs are non-Abelian anyons, implying that quantum computing can be realized by just braiding them. Stimulated by these elegant ideas, the pursuit of MFs in solid state systems has last for years \cite{Alicea}. So far, there already exist a lot of theoretical proposals, such as the $\nu=5/2$ fractional quantum Hall state \cite{Read}, $p$-wave superconductors \cite{Kitaev, Ivanov}, the surface states of topological insulator with proximity to an $s$-wave superconductor \cite{Fu}, one-dimensional semiconducting nanowires (SNW) \cite{Oppen}, and so on. Notably, elementary progress has been made in experiments on SNW most recently \cite{Mourik}, which makes the MF based topological computation really promising.

If the building blocks of MFs finally become true, and braiding operations are also achieved in this promising system of SNW, as predicted by Alicea et. al. \cite{Alicea2}, how to read out and initialize topological qubits will be another challenge towards the MF based topological quantum computation. The difficulty of detecting the topological qubits hides exactly in their robustness, for the detection itself is actually a special decoherence process. There already exist several proposals, including using the interferometry \cite{Bonderson}, fractional Josephson effects \cite{Fu2}, coupling with normal qubits \cite{Beenakker} or utilizing quantum dots (QDs) and a tunable flux \cite{Flensberg}. While these methods may play important roles during the exploratory stage, for the final purpose of large scale implementation, more practical and efficient proposals are still highly demanded.

Here we propose to read out and initialize a topological qubit in SNW by using single level QDs. We show that by coupling two MFs through gate voltage, which removes the parity degeneracy, the topological qubit can be directly measured through the occupation of the QD in an energy window. The initialization process can also be achieved by tuning the gate voltage of the QD.

\begin{figure}
\centering
\includegraphics[width=0.45\textwidth]{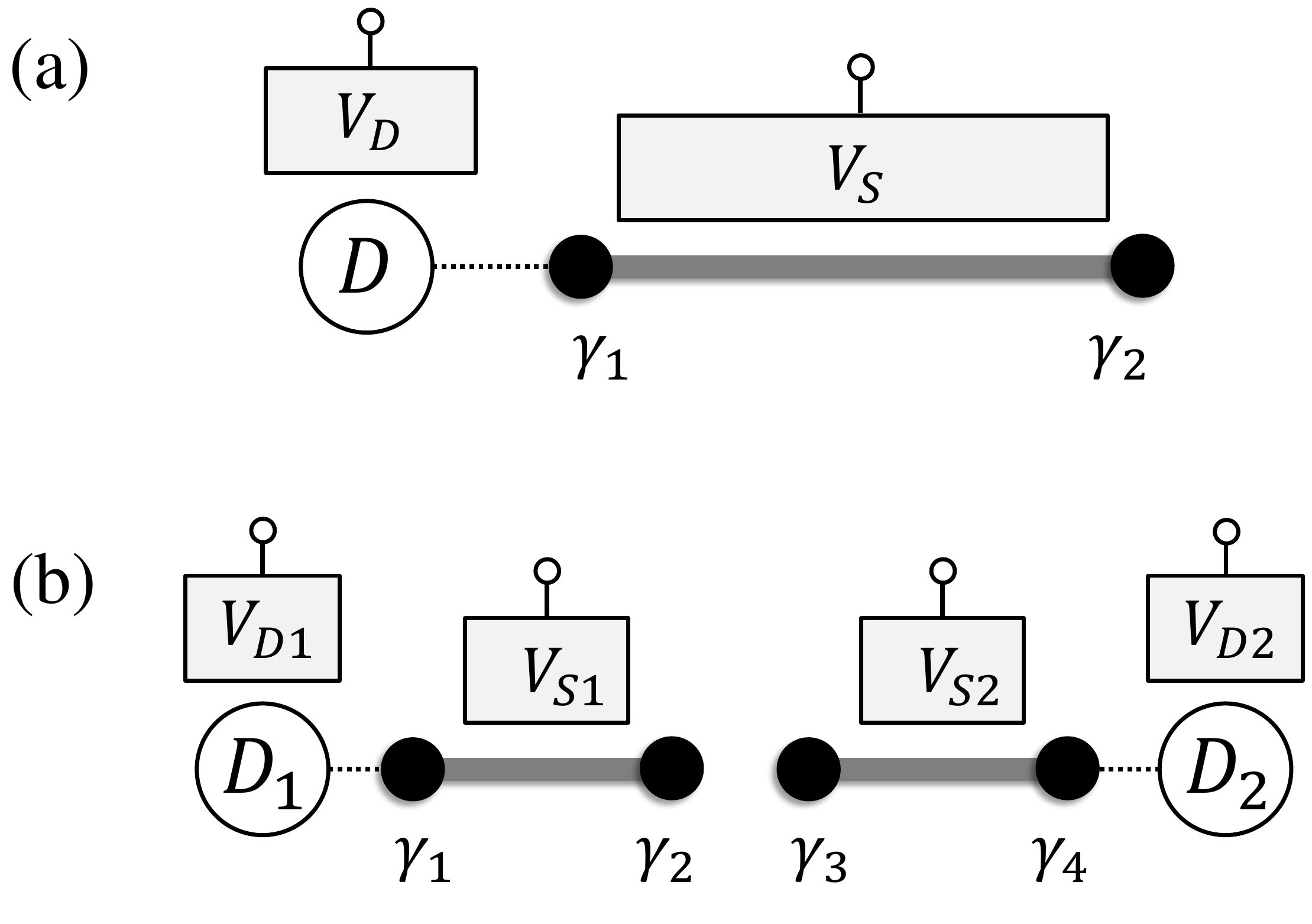}
\caption{Illustration of the proposed setup. (a) A semiconducting nanowire in the topological superconducting phase supports a pair of MFs $\gamma_{1,2}$ located at the ends. The gate voltage $V_S$ deposited on the nanowire can be used to adjust the coupling between two MFs. A QD $D$ is coupled to $\gamma_1$ on the left side, with its energy level tuned by the gate $V_D$. (b) A more practical structure composed by four MFs $\gamma_{1,\cdots,4}$, two QDs $D_{1,2}$ and also four gates $V_{S1,2}$, $V_{D2,2}$.} \label{fig1}
\end{figure}

We start from a simple case, where the topological qubit is carried by only one pair of MFs, as sketched in Fig. \ref{fig1}(a). An SNW is set into the nontrivial topological superconducting phase, where a pair of MFs $\gamma_1, \gamma_2$ are located at the ends of the wire. The MF is a neutral quasiparticle, which is its own anti-particle, or mathematically, $\gamma^\dag_{1,2}=\gamma_{1,2}$. A usual Dirac fermion $f$ can be defined by them through the relation $f=(\gamma_1+i\gamma_2)/2$. The gate voltage $V_S$ applied on the SNW can control the coupling strength between $\gamma_1$ and $\gamma_2$ \cite{Alicea2}, which can be modeled by an effective Hamiltonian \cite{Sarma}

\begin{equation}\label{hm}
H_M=i\frac{E_M}{2}\gamma_1\gamma_2,
\end{equation}
where $E_M$ is the quasiparticle excitation energy due to the coupling between the two MFs and a coefficient $1/2$ is from the particle-hole redundancy. In the absence of coupling between MFs, the excitation energy of quasiparticle is zero, which leads to degeneracy of the ground states. Such a coupling will lift the degeneracy, and it is an essential step of the detection of the topological qubit \cite{Fu2, Flensberg}.

A single level QD is weakly coupled to the Majorana bound state $\gamma_1$, which leads to electron tunneling between the QD and SNW. Note that the much weaker coupling to $\gamma_2$ can be dropped, without changing the main results. Such coupling between usual fermion and MF can be expressed by the tunneling Hamiltonian as \cite{Flensberg}
\begin{equation}\label{ht}
H_T=\Gamma(c^\dag-c)\gamma_1,
\end{equation}
where $\Gamma$ is the tunneling strength, which has been set real for simplicity, since its phase is not important here, and $c$ is the fermion operator of the QD.

Finally, the total Hamiltonian can be obtained by combining the unperturbed part of QD and Eq. (\ref{hm}, \ref{ht}),
\begin{equation}\label{h}
H=\varepsilon c^\dag c+\Gamma(c^\dag-c)(f+f^\dag)+E_M(f^\dag f-1/2),
\end{equation}
where the equalities $\gamma_1=f+f^\dag$ and $\gamma_2=i(f^\dag-f)$ have been used. The energy level $\varepsilon$ measured relative to the Fermi energy of the superconductor can be tuned by the gate $V_D$ (Fig. \ref{fig1}). We assume that all the energy scales in Eq. (\ref{h}) are lower than the superconducting gap $\Delta$ of the SNW, so the excitations above $\Delta$ can be neglected.

Though the coupling between the QD and SNW may change both of their own fermion parities, the total parity of the combined system remains conserved. Therefore when we write the combined Hamiltonian in a matrix form, the elements between different parities are absent. Specifically, the $2\times2$ matrices for even ($e$) and odd ($o$) parities can be expressed in a compressed form as
\begin{equation}\label{m}
H_{e,o}=\left(
          \begin{array}{cc}
            \mp E_M/2 & \Gamma \\
            \Gamma & \varepsilon \pm E_M/2 \\
          \end{array}
        \right),
\end{equation}
with the basis $\{|0\rangle_D|0\rangle_M, |1\rangle_D|1\rangle_M\}_e$ for even parity and $\{|0\rangle_D|1\rangle_M, |1\rangle_D|0\rangle_M\}_o$ for odd parity, respectively, while $|\cdots\rangle_{D,M}$ denote the fermion numbers in the QD and Majorana bound states.

The Hamiltonian in Eq. (\ref{m}) can be diagonalized directly. Taking the even parity case as an example, the excited state $|\psi_{e,+}\rangle=(\sin\frac{\theta_e}{2},\cos\frac{\theta_e}{2})$ and ground state $|\psi_{e,-}\rangle=(\cos\frac{\theta_e}{2},-\sin\frac{\theta_e}{2})$ are corresponding to the eigenvalues $E_e^{\pm}=\frac{\varepsilon}{2}\pm\sqrt{(\frac{\varepsilon+E_M}{2})^2+\Gamma^2}$, where $\theta_e=\cot^{-1}(\frac{\varepsilon+E_M}{2\Gamma})$. The results for the case of odd parity can be obtained by a simple substitution, $E_M\rightarrow -E_M$, and specifically, we have $\theta_o=\cot^{-1}(\frac{\varepsilon-E_M}{2\Gamma})$, and $E_o^{\pm}=\frac{\varepsilon}{2}\pm\sqrt{(\frac{\varepsilon-E_M}{2})^2+\Gamma^2}$.

We here focus on the time evolution of the initial states ($t=0$) under the variation of $\varepsilon(t)$ and $E_M(t)$, with $\Gamma$ being fixed all the time. The process can be done adiabatically \cite{Flensberg}, as long as the variation of the Hamiltonian is slowly, or more precisely, $\hbar\dot{\varepsilon}\ll \Gamma^2$ and $\hbar\dot{E}_M\ll \Gamma^2$. The initial state of topological qubit is assumed in a general form of $|\psi\rangle_M=\alpha|0\rangle_M+\beta|1\rangle_M$ with the normalization condition $|\alpha|^2+|\beta|^2=1$, and the state of the QD is set to $|0\rangle_D$ by tuning its energy level in the limit of $\frac{\varepsilon-E_M}{\Gamma}\gg1$. Then, we adiabatically turn on the gate voltages $V_{D,S}$, and correspondingly, the state will slowly evolve into
\begin{equation}\label{wave}
\begin{split}
|\psi(t)\rangle=\alpha e^{i\phi(t)} \big(\cos\frac{\theta_e}{2}|0\rangle_D|0\rangle_M-\sin\frac{\theta_e}{2}|1\rangle_D|1\rangle_M\big)\\
+\beta\big(\cos\frac{\theta_o}{2}|0\rangle_D|1\rangle_M-\sin\frac{\theta_o}{2}|1\rangle_D|0\rangle_M\big),
\end{split}
\end{equation}
with a relative phase $\phi(t)$ accumulated during the evolution. The average fermion nunmber in the QD can be calculated with respect to the ground state in Eq. (\ref{wave}) as
\begin{equation}\label{n}
\langle n_D\rangle=|\alpha|^2\sin^2\frac{\theta_e}{2}+|\beta|^2\sin^2\frac{\theta_o}{2}.
\end{equation}

\begin{figure}
\centering
\includegraphics[width=0.45\textwidth]{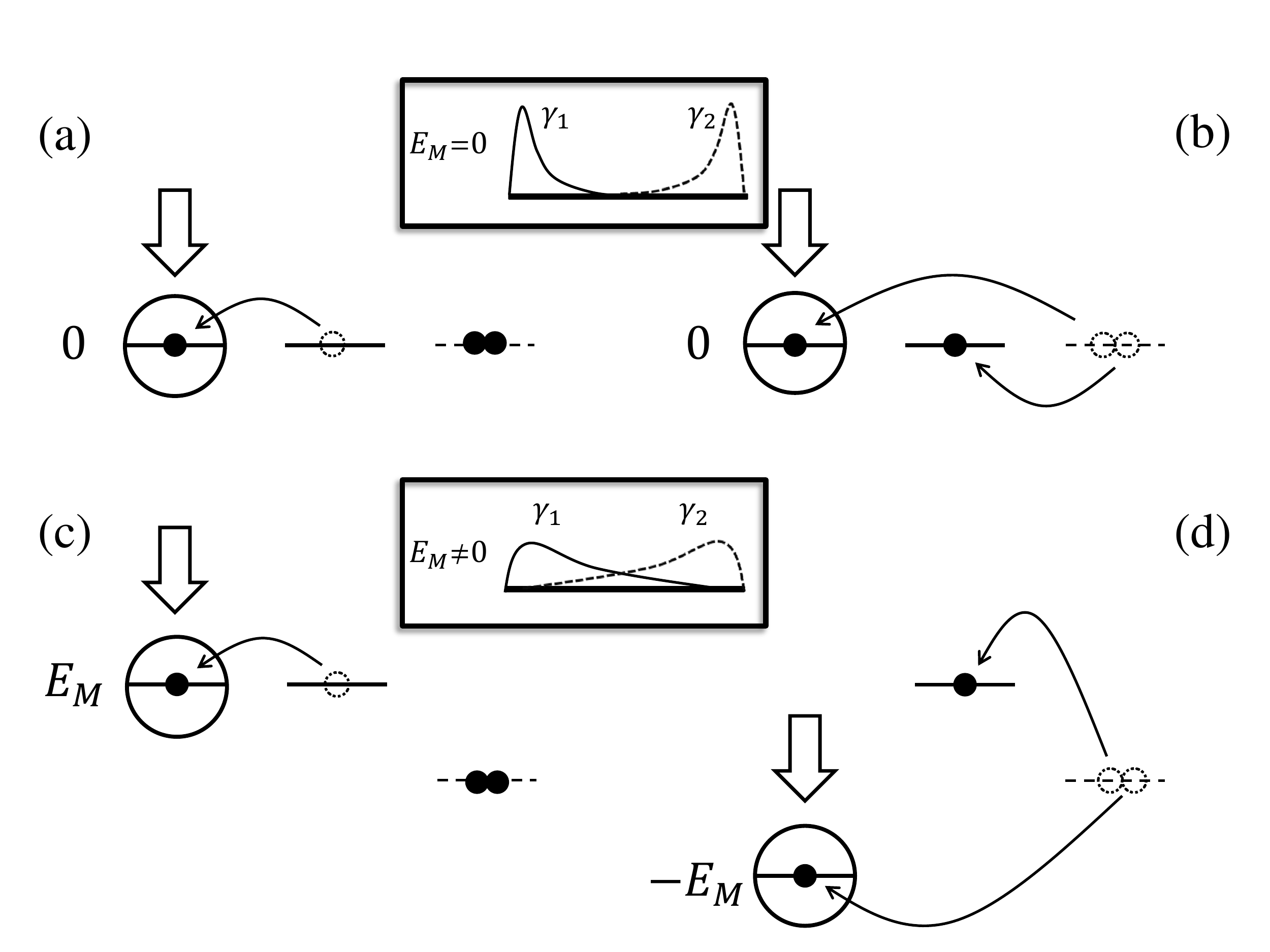}
\caption{Reading out a topological qubit by lowering the energy level of the QD. The circles denote QDs, and the black dots are electrons. The solid level is the fermion energy composed by two MFs, while the dashed level stand for the condensate of the superconductor, which is filled by Cooper pairs. (a, b) No coupling exists between MFs (as the upper inserted box), so the ground states are parity degenerate. When the energy level is near chemical potential $0$ of the SNW, there is an electron jump into QD, no matter the initial state is $|1\rangle_M$ or $|0\rangle_M$. (c, d) Coupling between MFs is introduced by gate voltage (as the lower inserted box), and the ground state degeneracy is lifted. The change of electron number in the QD occurs only when the level of QD reaches $E_M$ and $-E_M$, corresponding to the initial states of $|1\rangle_M$ and $|0\rangle_M$, respectively.} \label{fig2}
\end{figure}

This expression indicates that the topological qubit can be measured by the occupation of the QD. To see this point clearly, we first investigate the limiting case of $|\alpha|=0$, i.e., a single quasiparticle occupies the Majorana bound states initially. We fix $E_M$ to a positive value, and then tune the energy level of the QD downward by the gate from $\frac{\varepsilon-E_M}{\Gamma}\gg1$ to $\frac{\varepsilon-E_M}{\Gamma}\ll-1$, as sketched by Fig. \ref{fig2}. At first, the occupation of QD remains unchanged, until $\varepsilon$ approaches $E_M$, where the electron in the Majorana bound states starts to jump into the QD as in Fig. \ref{fig2}(a, c), in order to lower the total energy. If we continue to lower the energy level to the region $\frac{\varepsilon-E_M}{\Gamma}\ll-1$, the occupation of the QD will be nearly $1$. In the opposite limit $|\alpha|=1$, i.e., there is no quasiparticle occupying the Majorana bound states initially, the results may be different. Since there is no electron in the excitation level $E_M$, when the energy level $\varepsilon$ passes by, its occupation will not change. However, if we continue to lower the level of the QD, when it approaches $-E_M$, one electron in SNW will then transfer into the QD, as shown by Fig. \ref{fig2}(b, d). This result is not surprising if we take the condensate of SNW into consideration as well. When the QD level reaches $-E_M$, the Cooper pair at the chemical potential of the superconducting SNW will be splitted into two excitation quasiparticles, one in the excitation state formed by MFs and the other tunneling into the QD as in Fig. \ref{fig2}(b, d), for such a state is energetically favorable. As the former case, in the limiting condition $\frac{\varepsilon+E_M}{\Gamma}\ll-1$, the fermion number in QD is also nearly $1$.

\begin{figure}
\centering
\includegraphics[width=0.45\textwidth]{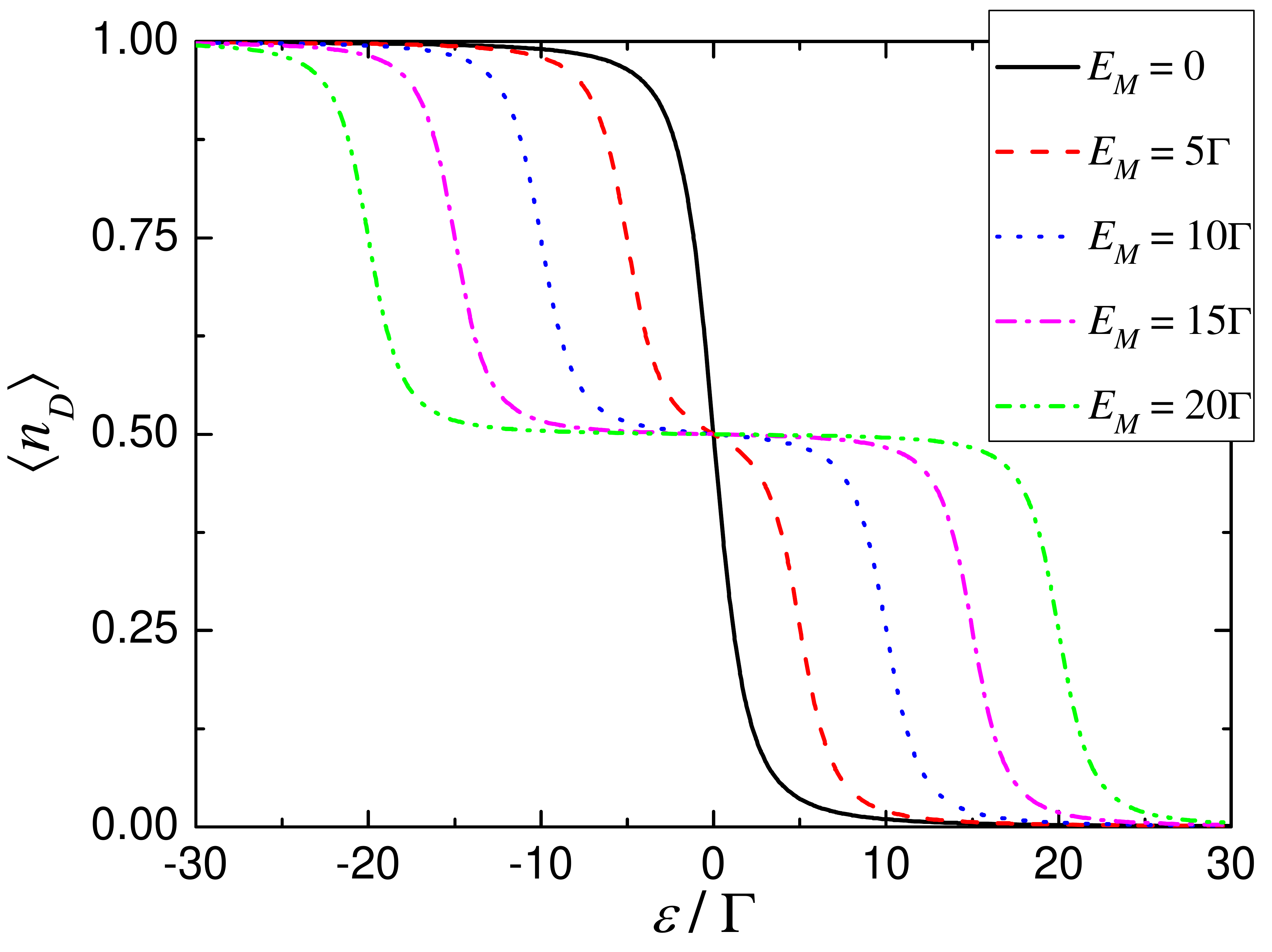}
\caption{(color online) Plot of the occupation number of the QD $\langle n_D\rangle$ as a function of $\varepsilon$, with different values of $E_M$. The state of the qubit is set as $|\alpha|^2=0.5$.} \label{fig3}
\end{figure}

\begin{figure}
\centering
\includegraphics[width=0.45\textwidth]{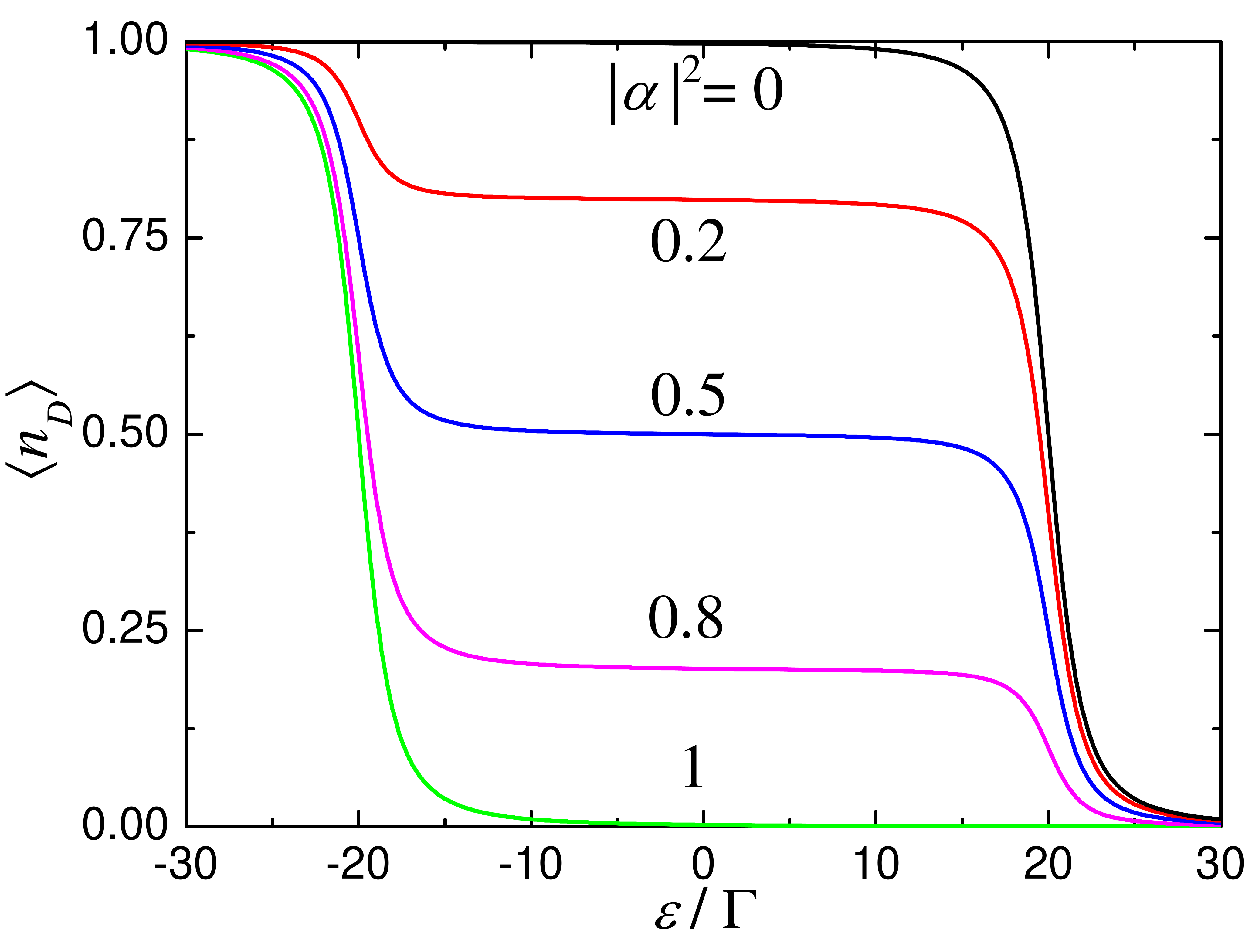}
\caption{(color online) Plot of the occupation number of the QD $\langle n_D\rangle$ as a function of $\varepsilon$, with different initial states of the topological qubit, where the coupling strength is chosen as $E_M=20\Gamma$.} \label{fig4}
\end{figure}

From above discussions we can draw an important conclusion that there exists an energy window $-E_M+\Gamma\ll\varepsilon\ll E_M-\Gamma$ for the QD, where the fermion number in it is $1$ for state $|1\rangle_M$ and $0$ for state $|0\rangle_M$, or more generally, $|\beta|^2$ for $|\psi\rangle_M=\alpha|0\rangle_M+\beta|1\rangle_M$. This property can be utilized to read out the topological qubit. However, to achieve this goal, the condition $E_M\gg\Gamma$ should be fulfilled. Physically, this means the broadening of the QD level characterized by $\Gamma$ is much less than the energy splitting $E_M$ due to the coupling between the MFs. Therefor, the coupling between the MFs is of great importance in the present proposal, which makes the original parity degenerate states now distinguishable by their energies. Without such coupling, the energy window is closed, so the fermion number in QD is always the same for both $|1\rangle_M$ and $|0\rangle_M$ at all energy scales. All the discussions and results mentioned above can be understood more clearly with the help of the schematic tunneling picture in Fig. \ref{fig2}.

Besides the above qualitative conclusion, we present also some numerical results. The average fermion number $\langle n_D\rangle$ in QD as a function of $\varepsilon$ is shown in Fig. \ref{fig3} and \ref{fig4}. In Fig. \ref{fig3}, different coupling intensities $E_M$ between MFs are enumerated with the same qubit state as $|\alpha|^2=0.5$. It is clear that there is an platform $\langle n_D\rangle\simeq0.5$ within the energy window $(-E_M,E_M)$, which can be utilized to read out the topological qubit. As the window gets broader, the measurement becomes more precise. In the uncoupling limit $E_M=0$, the energy window is completely closed, as is the case of parity degeneracy. The results are shown for different initial states of the topological qubit with $E_M=20\Gamma$ in Fig. \ref{fig4}. The heights of the platform are equal to $|\beta|^2$.

So far, we have focused on the topological qubit composed by two MFs. Given that braiding operators preserve the fermion parity and the qubit state $|\psi\rangle_M=\alpha|0\rangle_M+\beta|1\rangle_M$ is a superposition of different parity states, and thus such a qubit is not actually usable. Alternatively, we may utilize four MFs to encode a topological qubit \cite{Ivanov, Bravyi}, as $|\psi\rangle'_M=\alpha|00\rangle_M+\beta|11\rangle_M$, where $|00\rangle_M$ and $|11\rangle_M$ are the eigenstates of two Dirac fermions with the former being defined by $\gamma_{1,2}$ and the latter by $\gamma_{3,4}$, as shown in Fig. \ref{fig1}(b). Due to parity conservation, quantum computation is performed in the same parity subspace of two Dirac fermions. As the occupation number of both Dirac fermions is always the same, the qubit can be read out either by tuning $D_1$ or $D_2$ (Fig. \ref{fig1}(b)), by utilizing the same approach as discussed for the illustrative qubit $|\psi\rangle_M$.

Finally we would like to show that the initialization of the topological qubit can also be achieved by adjusting gate voltages of QDs. First we note that the topological qubit can not be initialized without knowing its original state. The reason is just the same as that we can not read out the qubit when the parity degeneracy occurs. One can always change the fermion number in the Majorana bound states, regardless of even or odd parity. Thus the initialization process can be achieved based on the results already read out. Specifically, if the read out result is $|00\rangle_M$, then we can initialize the qubit to $|00\rangle_M$ ($|11\rangle_M$) by tune the energy levels of the QDs $D_{1,2}$ to the limit of $\frac{\varepsilon-E_M}{\Gamma}\gg1$ ($\frac{\varepsilon+E_M}{\Gamma}\ll-1$). In the same spirit, if the result is $|11\rangle_M$, then the qubit can be initialized to $|00\rangle_M$ ($|11\rangle_M$) by setting the energy level of QDs as $\frac{\varepsilon+E_M}{\Gamma}\ll-1$ ($\frac{\varepsilon-E_M}{\Gamma}\gg1$). It should be noted that the fermion parity before and after the reading out process is always opposite. After the initialization, coupling between MFs should be removed, to regain the topological protection.

In summery, we present a proposal to read out and to initialize topological qubits by QDs. Fusing the MFs first by gate voltage, then the qubit is measured through the occupation number of the QDs in the energy window decided by the coupling strength between the MFs. The initialization of the qubit is achieved by a following operation on QDs using gate voltages. Both processes utilize the gate voltages only, in an all-electrical way.

\begin{acknowledgments}
This work was supported by 973 Program (Grants No. 2011CB922100, No. 2011CBA00205, and No. 2013CB921804), by NSFC (Grants No. 11074111, No. 11174125, No. 11023002 and No. 11004065), by PAPD of Jiangsu Higher Education Institutions, by NCET, by the PCSIRT, by the Fundamental Research Funds for the Central Universities, by the GRF (HKU7058/11P) and CRF (HKU-8/11G) of the RGC of Hong Kong.
\end{acknowledgments}

\end{document}